\documentclass[fleqn,10pt]{wlscirep}
\usepackage[utf8]{inputenc}
\usepackage[T1]{fontenc}
\title{\fontsize{18}{18}\selectfont Queue \& AI: When Faster Tasks Slow Down the Workflow}

\usepackage{comment}
\usepackage{verbatim}
\usepackage{mathtools}
\usepackage{graphicx}
\usepackage{hyperref}
\usepackage{xcolor}
\usepackage[font=small,labelfont=bf]{caption}
\usepackage{parskip}

\newcommand{\E}{\mathbb{E}}
\newcommand{\Var}{\mathrm{Var}}

\usepackage{float}
\author[1,]{Silvia Bartolucci}
\author[2,*]{Pierpaolo Vivo}
\affil[1]{Department of Computer Science, University College London, United Kingdom}
\affil[2]{Department of Mathematics, King’s College London, United Kingdom}

\affil[*]{pierpaolo.vivo@kcl.ac.uk}

\begin{abstract}
\noindent Quantifying the workplace productivity effects of Generative Artificial Intelligence is now central to economics, management, and public policy. The deployment of AI tools in customer service, writing, software development, and consulting operations has been reported to generate large per-task productivity gains, typically measured as tasks completed per worker-hour or reductions in mean handle time. We argue that such mean-based metrics can misrepresent AI’s effects in workflows where tasks accumulate and compete for scarce human attention.  AI assistance can generate a deceptive productivity signature: average completion times fall because AI tools typically supply a fast first draft, yet workflow-level performance deteriorates when a  subset of AI errors escapes review and returns as costly downstream rework. We call this divergence between mean task speed and system-level delay the \textit{variance wedge}.
Depending on the operational parameters, the most time-efficient way to complete a workflow may undergo a transition between two task-processing regimes, a fully AI-assisted and a fully manual one. We formalize the mechanism as a queueing model and derive two main implications analytically. First, under congestion, reviewers rationally raise the risk threshold for checking AI outputs, reducing scrutiny precisely when it would matter the most. Second, AI assistance can stabilize an overloaded workflow only when (i) the fraction of tasks handled by AI exceeds a critical threshold, and (ii) the human attention required for review and expected rework is lower than the attention for manual completion, a requirement substantially more stringent than faster draft generation. These results suggest that AI deployment should be evaluated not only by average task speed, but by its overall effects on congestion, rework, and the robustness of human oversight under load.
\end{abstract}


\begin{document}

\maketitle

\section{Introduction}

Consider a software team at a large company after the adoption of an Artificial Intelligence (AI) coding assistant. The first quarterly report looks encouraging. Developers generate more code, publish new features, and complete more tickets. At the upstream stage of the workflow, AI appears to do exactly what it promises: it turns natural-language prompts into working-looking code at low human cost and considerably higher speed. But software development is not finished when the code is generated. Code must be reviewed, tested, integrated, deployed, maintained, and sometimes repaired after failure. A second report, six months later, may therefore tell a different story. Bug-fix requests grow larger and review queues lengthen as more defects are discovered downstream. Incidents rise, and senior engineers spend more time reading, correcting, and maintaining code that looked perfectly functional at first glance. In this process, the productivity bottleneck has  moved from generation to verification, a shift that business productivity models and forecasts of economic growth must account for. 

Recent evidence on AI-assisted coding supports this new operational model with empirical data. Ref. \cite {daniotti2026using} documents the diffusion of AI-generated code on GitHub, showing that generative AI already shapes production and code review at scale. Ref. \cite{wu2026ai} further argues that productivity gains are achieved when workers with enough expertise are available to judge, adapt, and maintain AI outputs, with AI effectively raising the productivity bar. Ref. \cite{BurnMurdochOConnor2026} shows that in organizations self-reported individual productivity gains are systematically overstated, and that the technology accelerates individual upstream work, while simultaneously loading the downstream stages where the organizational output is actually produced.

The picture these findings delineate is consistent: AI accelerates generation, but generation is not the binding productivity constraint. The constraint is human attention necessary to review, integrate, and respond to incidents:  AI  loads the human capabilities with both more new outputs to inspect and a heavier tail of outputs that fail closer scrutiny downstream (e.g., when AI hallucinates and produces very plausible but incorrect outputs \cite{ji2023,huang2025}). The relevant productivity question is, therefore, not whether AI makes generation faster, but whether the full workflow is. This shift\cite{wu2026ai} is common to similar task completion workflows in organizations belonging to very different business sectors, from customer service \cite{brynjolfsson2025} and professional writing \cite{noy2023} to consulting \cite{dellacqua2026}.

Quantifying the productivity impact of generative AI has become a central problem for economics, management, and public policy to drive investment decisions, shape regulations, make labor-market forecasts, or improve organizational design. 
In a nationally representative U.S. survey, 28\% of workers reported using generative AI at work, and recent users reported that it assisted a non-trivial share of weekly work hours \cite{BickBlandinDeming2025}. Field experiments and large-scale observational studies document substantial average gains in customer service, professional writing, software development, and consulting \cite{brynjolfsson2025,noy2023,peng2023,dellacqua2026}. A recent review \cite{del2025ai} summarizes the theoretical and empirical evidence on the impact of AI productivity on skills and jobs, reporting productivity gains on the order of 20--60\% in randomized trials and 15--30\% in field experiments, with significant heterogeneity by skill level and task complexity. 

These studies typically measure mean outcomes: tasks completed per worker-hour, mean handle time, completion speed, or average quality at a fixed time. These measurements have shaped managerial deployment decisions, and increasingly serve as inputs into macroeconomic projections of AI-driven growth. In task-based accounting frameworks such as Acemoglu's productivity model \cite{acemoglu2025simple}, aggregate productivity effects on GDP growth are obtained by combining the share of tasks exposed to AI with average task-level cost savings.
This approach does not tackle the corresponding microeconomic question central to organizations deploying AI: \textit{how do task-level time savings translate into workflow-level performance when work arrives over time, competes for scarce human attention, and sometimes returns as rework?} 

\begin{figure}[H]
    \centering
    \includegraphics[width=0.9\linewidth]{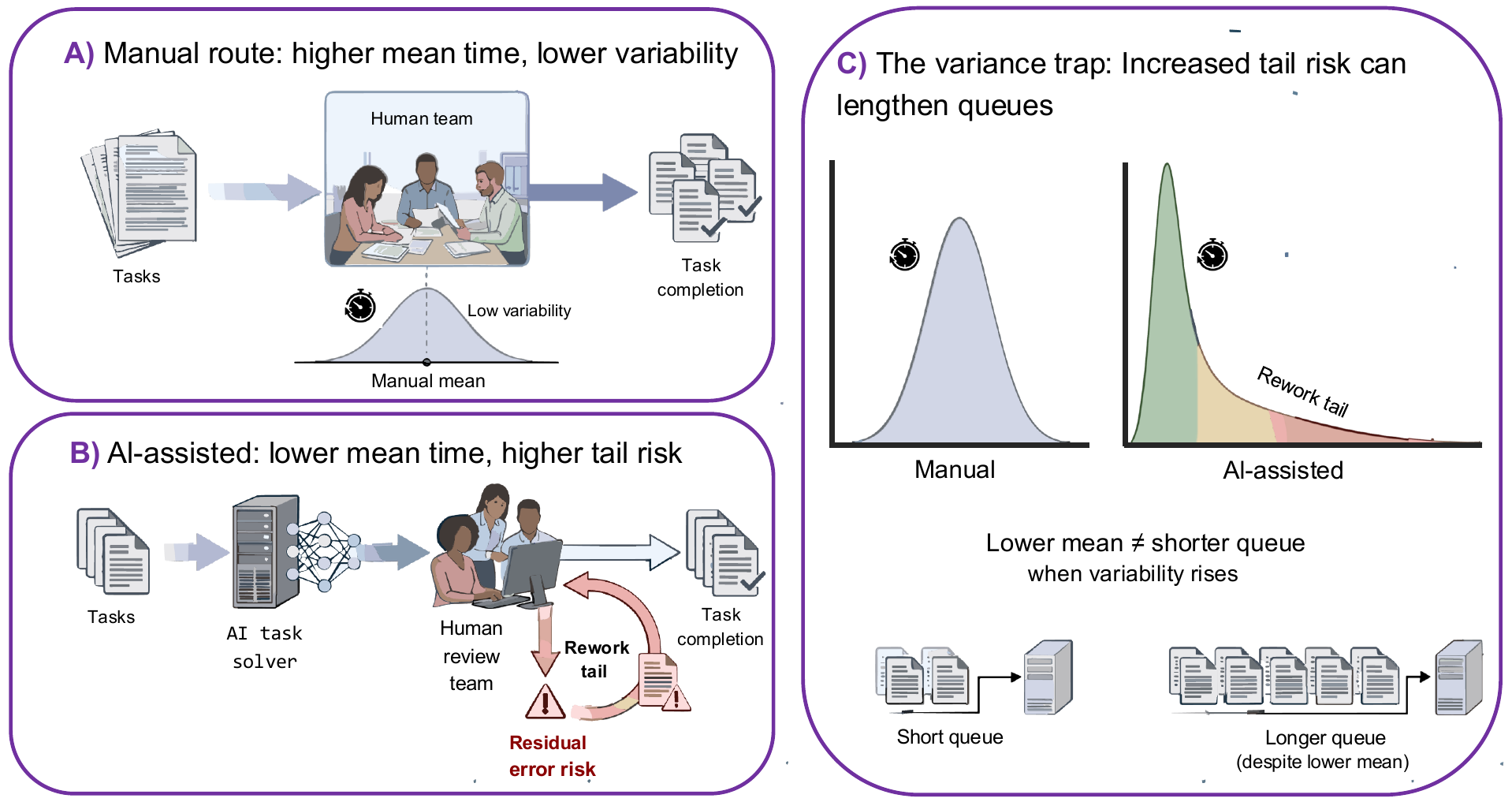}
    \caption{Schematic workflow model. Panel A shows the manual route: tasks require more human time on average but have relatively low variability. Panel B shows the AI-assisted route: AI generates a draft cheaply, but human review is imperfect, so some residual errors escape and return as rework. Panel C illustrates the resulting distributional shift. AI lowers the mean service time, but creates a heavier right tail. As queueing delay depends on the second moment of service time, the rework tail may dominate the mean saving, producing longer queues despite faster average task completion. }
    \label{fig:scheme}
\end{figure}

This paper proposes a queueing model in which tasks arrive randomly to a human team and are handled either manually or with AI assistance. Under manual handling, the team completes the task directly. Under AI assistance, the AI generates an answer, a code snippet, or a recommendation cheaply. A human reviewer spends time checking it. Some outputs that pass review are operational false positives: they appear functional at handoff, but later fail in deployment, customer interaction, audit, legal scrutiny, or integration with the rest of the workflow. When such failures return as rework, AI-assisted service time -- defined as the total amount of human attention it consumes, including review and any later correction work -- acquires a distinctive shape, characterized by a lower mean but a heavier  tail. In Fig.\ref{fig:scheme} we report a schematic representation of the different workflows under the manual or AI-assisted route. Using the Pollaczek–Khinchine and Kingman frameworks developed in queueing theory \cite{pollaczek1930, khinchine1932, kingman1961} we derive a closed-form expression for a relevant productivity metric affected by the overall service time distribution, namely the average time 
$W_q$ a task has to wait in the queue before being attended to (see Eq. \eqref{eq:Wq_model}). From this formula, we derive an analytical condition for what we call the \emph{variance wedge effect} (see Eq. \eqref{eq:trap_CV}): AI reduces mean human time per task, while increasing mean waiting time across the workflow. Since many economically important uses of generative AI (customer-support cases, legal reviews, medical triage decisions, insurance claims, document adjudication, and content moderation) involve tasks that line up for scarce human attention, we argue that classical queueing theory provides a relevant analytical framework to address these questions and shows that these two metrics--average time consumed by tasks, and mean waiting time across the workflow--need not necessarily move together.

The model also yields two implications about human review and system stability. First, when reviewer attention becomes scarce, the rational response is not necessarily to check AI outputs more carefully. Since review time competes with all other tasks waiting for human attention, congestion actually raises the cost of scrutinizing any individual draft. The optimal checking rule, therefore, becomes more selective: the reviewer raises the risk threshold above which an AI output is examined, passes more low- and intermediate-risk drafts through unchecked, and reduces review effort even for drafts whose perceived risk remains high enough to warrant closer inspection. Second, AI can stabilize an overloaded workflow only if it genuinely reduces the total human attention demand imposed on the system. This requires both sufficient adoption (the fraction of AI-routed tasks must exceed a critical threshold, see Eq. \eqref{eq:xc}), and a favorable accounting of the AI route itself: review time plus the expected rework generated by escaped errors must be lower than the human time required for manual handling. Thus faster generation alone is not the relevant criterion: an AI system may produce drafts quickly while still failing to relieve the queue if those drafts require substantial checking or create enough downstream correction work.

The rework mechanism we study is related to the classical theory of feedback queues. In a queue with Bernoulli feedback, a customer who completes service leaves the system with probability \(1-p\), but with probability \(p\) returns to the back of the same queue for another service attempt. This class of models goes back at least to Takács' single-server queue model with feedback \cite{takacs1963single}, and was later developed in several works  \cite{disney1980mg1,disney1981sojourn,choi2000mg1}. Feedback is also a standard feature of queueing networks: in open Jackson networks, a job leaving one station may be routed to another station, or back to a previously visited station, thereby generating endogenous traffic in addition to the external arrival stream \cite{jackson1957networks,kleinrock1976queueing}. In our model, rework precisely acts as a feedback stream: it is traffic generated by past service completions, and can amplify congestion even when the exogenous task arrival rate is unchanged. Very recently, queueing-theoretic methods have also been applied to study congestion inside the serving infrastructure of deployed Large Language Models (LLM). For example, Ref. \cite{yang2024queueing} model LLM inference as an \(M/G/1\) queue in which variable output lengths drive service-time variability, while Ref. \cite{ozbas2026queueing} optimize task-specific reasoning-token budgets by trading off accuracy gains against the queueing delays induced by longer inference times. Ref. \cite{mitzenmacher2025queueing2} surveys how predictions of service time, request size, and reasoning complexity can improve scheduling in LLM inference systems, where latency is shaped by variable job sizes, batching, preemption, and memory constraints.

Throughout the paper, we use ``AI'' in an implementation-agnostic sense. The model does not depend on
a particular architecture, vendor system, benchmark score, prompting protocol, or
software-engineering choice. What matters is that the system has three features:
it reduces the human time needed to produce a first version of the output; it
leaves some residual probability of error after human review; and errors that
pass review can create later correction work. This includes, but is not limited
to, systems that generate fluent but unreliable text, badly written or sloppy code, or
questionable but very confident recommendations (what is sometimes colloquially called \emph{AI slop}). In this context, AI modifies typical quality signals (e.g., clarity of writing style) and raises the bar to identify high quality work \cite{kusumegi2025scientific}. The benefit of
this abstraction is that the analysis is not tied to a specific model family or
implementation. The limitation is that benchmark results do not necessarily map one-to-one
onto the model parameters. A benchmark may measure task accuracy, factuality, or
coding performance in isolation, whereas the quantities in the model are
workflow-level objects: review time, residual error probability after review, and
the rework generated by errors that escape. Generative AI is therefore not the only possible source of this service-time pattern, but
it is currently the most prominent and empirically important one.

Studies of AI productivity in workflow settings should therefore report the full distribution of service times, including downstream rework, not only the mean. The implication for policy and measurement is that aggregate productivity statistics combining task-level exposure with mean cost savings (the standard input to macroeconomic accounting of AI) can systematically overstate organizational gains when rework reallocates rather than removes work. The model thus supplies a microeconomic foundation for macro frameworks of AI productivity, identifying the second-moment and rework terms that mean-based aggregation omits. The deeper point is that AI adoption can shift a workflow between regimes, in a sense familiar from traffic flow: a road may carry cars smoothly until density crosses a critical threshold, after which interactions among vehicles produce a sharp collapse in average speed. The slowdown is not caused by any one car becoming intrinsically slower, but by the collective effect of operating too close to capacity. AI-assisted workflows admit a similar transition. The mean gain on the typical task can coexist with, and ultimately produce, a congested regime in which overall waiting times rise sharply. Mean productivity statistics, computed within the smooth regime, miss this important effect entirely.

\section{Results}

\subsection{The model}\label{sec:modelmain}

We model an organization as a single-server queueing system in which tasks compete for scarce human attention. A task is a unit of work that must ultimately be cleared by the human review system: for example, a customer-support case, a legal file, a claim, a code change, or a moderation decision. Tasks arrive at rate \(\lambda>0\) and are routed either to full manual handling or to AI-assisted handling with probability \(1-x\) and \(x\), respectively, where \(x\in[0,1]\) is the AI-routing fraction. In the manual route, the task is handled entirely by humans. In the AI-assisted route, the AI supplies an initial draft, recommendation, or code snippet at negligible human cost, but the task still requires human review and may later return as rework if an error escapes.

The routing decision is taken \textit{ex ante}, before any task-specific signal is observed. The server processes work at a rate $C$, measured in hours of human attention per unit of calendar time.  Treating $C$ as a single aggregate team capability is a simplification: the model does not distinguish between sub-teams, seniority levels, or functional roles within the organization, even though in practice the binding constraint is typically narrower, often related to the availability of senior reviewers, expert adjudicators, or a small group of specialists who handle escalations. We interpret $C$ as the capacity of that scarce reviewing layer rather than as headcount or total labor hours, since it is that layer's attention that AI-assisted workflows load most heavily.

A manual task consumes a random amount $T_H$ of human-attention time with mean $\tau_H$ and variance $\sigma_H^2$. Its second raw moment is therefore $q_H:=\mathbb{E}[T_H^2]
=
\tau_H^2+\sigma_H^2
=
\tau_H^2(1+c_H^2)$, where we write the squared \emph{coefficient of variation} (CV)--a standardized measure of dispersion--as $c_H^2=\sigma_H^2/\tau_H^2$. No further structure is imposed on the manual task processing.

An AI-routed task proceeds in two stages. The AI generates a draft at negligible human cost, after which the reviewer spends $r\geq 0$ hours checking it. Review reduces, but does not eliminate, the probability that the draft contains an undetected error. Let $p(r)$ denote the residual miss probability after the review effort $r$. In the following, we use the following form for the probability:
\begin{equation}
p(r)=p_\infty+(p_0-p_\infty)e^{-\kappa r}\ .
\label{eq:p_r}
\end{equation}
This form captures three assumptions: review reduces errors, each additional unit of review has a smaller marginal effect, and some errors may be irreducible by review alone. The parameter $\kappa>0$ represents the reviewer's \emph{verification skill}: a higher $\kappa$ means each hour of review catches a larger fraction of errors. 

If an error escapes review, the task may return to the same human team through correction, escalation, complaint handling, audit, or downstream repair. The total human-attention cost of an AI-routed task is therefore
\begin{equation}
\label{eq:T_A}
T_A=r+M R\ ,
\end{equation}
where the random variable $M=1$ if the error escapes review and generates additional work, and $M=0$ otherwise; the random variable $R$ with $\mathbb E[R]=\mu_R$ and $\mathbb E[R^2]=\mu_{R,2}$ is the resulting additional human-attention requirement.

The unconditional service time $T$ for a randomly selected task is therefore $T_H$ with probability $1-x$ and $T_A$ with probability $x$. Its mean, denoted $m(x;r)$, is the average amount of human-attention time required by each incoming task. We also define as $q(x;r)$ the second moment (formulas for our setting are provided in Section \ref{sec:MM_model_primitives} below). Therefore, the mixed-system squared CV, which measures service time dispersion around the mean and is a proxy for how unpredictable the waiting time in the queue is, has the form $c_s^2(x;r)=q(x;r)/m(x;r)^2-1$.

Since tasks arrive at rate $\lambda$, the system receives on average $\lambda m(x;r)$ units of human-attention demand per unit of calendar time. The reviewing team can supply $C$ units of human-attention capacity per unit of calendar time. Hence, the utilization rate is $\rho(x;r)=\lambda m(x;r)/C$. For the queue to be stable, the average demand must be strictly smaller than the team capacity, so $\rho(x;r)<1$. If $\rho(x;r)\geq 1$, work arrives at least as fast as the team can process it, and waiting times grow without bounds.

For each task, we distinguish two time intervals. The first is the time the task spends waiting before the team starts working on it. The second is the service time \(S=T/C\), the time during which the team is actually occupied by performing that task. We denote the average waiting time before service begins by \(W_q\), which thus excludes the task's own service time; the average total time spent by the task in the system would be \(W_q+\mathbb E[S]\equiv W_q+m(x;r)/C\).

Our model thus defines a so-called $M/G/1$ queue. Therefore, under the stability condition \(\rho<1\), the Pollaczek--Khinchine formula \cite{pollaczek1930,khinchine1932} applies (see Section \ref{sec:PKsetting}), which gives for $W_q$ in our setting the expression
\begin{equation}
\label{eq:Wq_model}
\boxed{W_q(x; r) \;=\; \frac{\lambda\, q(x; r)}{2 C \,(C - \lambda m(x; r))}=\; \frac{\rho(x;r)}{1-\rho(x;r)}
\cdot
\frac{1+c_s^2(x;r)}{2}
\cdot
\frac{m(x;r)}{C}}\ .
\end{equation}
Eq. \eqref{eq:Wq_model} is our central result, from which all other implications follow. It shows that the waiting time is affected by three easily interpretable factors: a congestion factor \(\rho/(1-\rho)\), a variability factor \((1+c_s^2)/2\), and the mean calendar service time \(m/C\). Increasing the AI routing share \(x\) therefore changes waiting time not only by changing the average human time per task, but also by changing service-time variability, and by moving the system closer to, or farther from, the heavy-traffic region where the congestion multiplier becomes large.

For most of the paper, we assume Poisson task arrivals; the theory, however, can be easily adapted to bursty and irregular arrivals via the Kingman approximation \cite{kingman1961} described in \ref{sec:Kingman}, which simply amounts to a re-definition of the variability factor $(1+c_s^2)/2$ in Eq. \eqref{eq:Wq_model}. We will also define $W_A:=W_q(1;r)$ and $W_H:=W_q(0;r)$.

\subsection{The variance wedge}\label{sec:varwedge}

To evaluate the main tradeoff, we compare two extreme handling systems at the same arrival rate $\lambda$: a pure manual system ($x = 0$) and a pure AI-routed system ($x = 1$), holding the AI review effort $r$ fixed at some operational level. For readability, we write $\tau_A:=m(1;r)$, $q_A:=q(1;r)$ and $c_A^2:=c_A^2(r)=q_A/{\tau_A}^2-1$ to denote the average amount of human-attention time
required by each incoming task, its second moment, and the squared CV for a purely AI workflow, respectively.  We assume that both pure systems are stable: $\lambda\tau_H<C$ and $\lambda \tau_A<C$. 

Specializing \eqref{eq:Wq_model} to the $x=0,1$ cases, AI yields shorter mean waiting time than manual handling ($W_A<W_H$) if and only if
\begin{equation}
\label{eq:trap_CV}
\boxed{\;\;\frac{1 + c_A^2}{1 + c_H^2} \;<\; \left(\frac{\tau_H}{\tau_A}\right)^{\!2} \cdot \frac{1 - \lambda \tau_A/ C}{1 - \lambda \tau_H / C}\;\;}
\end{equation}
(see Section \ref{app:variancetrap} for a full derivation). This is the first main consequence that can be drawn from Eq. \eqref{eq:Wq_model}.

The left-hand side compares the spread of AI service times with the spread of manual service times. A value larger than one means that AI-routed tasks are less predictable: some may finish quickly after a short review, while others may return with costly rework. The right-hand side is the amount of extra unpredictability that the queue can absorb before AI stops being beneficial. We call it the AI \emph{variance budget}. When AI saves a lot of average time, this budget is large: the queue can tolerate occasional long rework cases. When AI saves only a little average time, the budget is small: even a thin tail of slow cases can erase the gains and lengthen the queue.

Two effects make this variance budget larger. The first is the most obvious direct time saving, captured by $\left(\tau_H/\tau_A\right)^2$. This term is greater than one whenever AI uses less human attention on average than manual work, \(\tau_A<\tau_H\). The more AI reduces the average task time, the larger this term becomes.
\begin{figure}[htb!]
\centering
\includegraphics[width=0.7\textwidth]{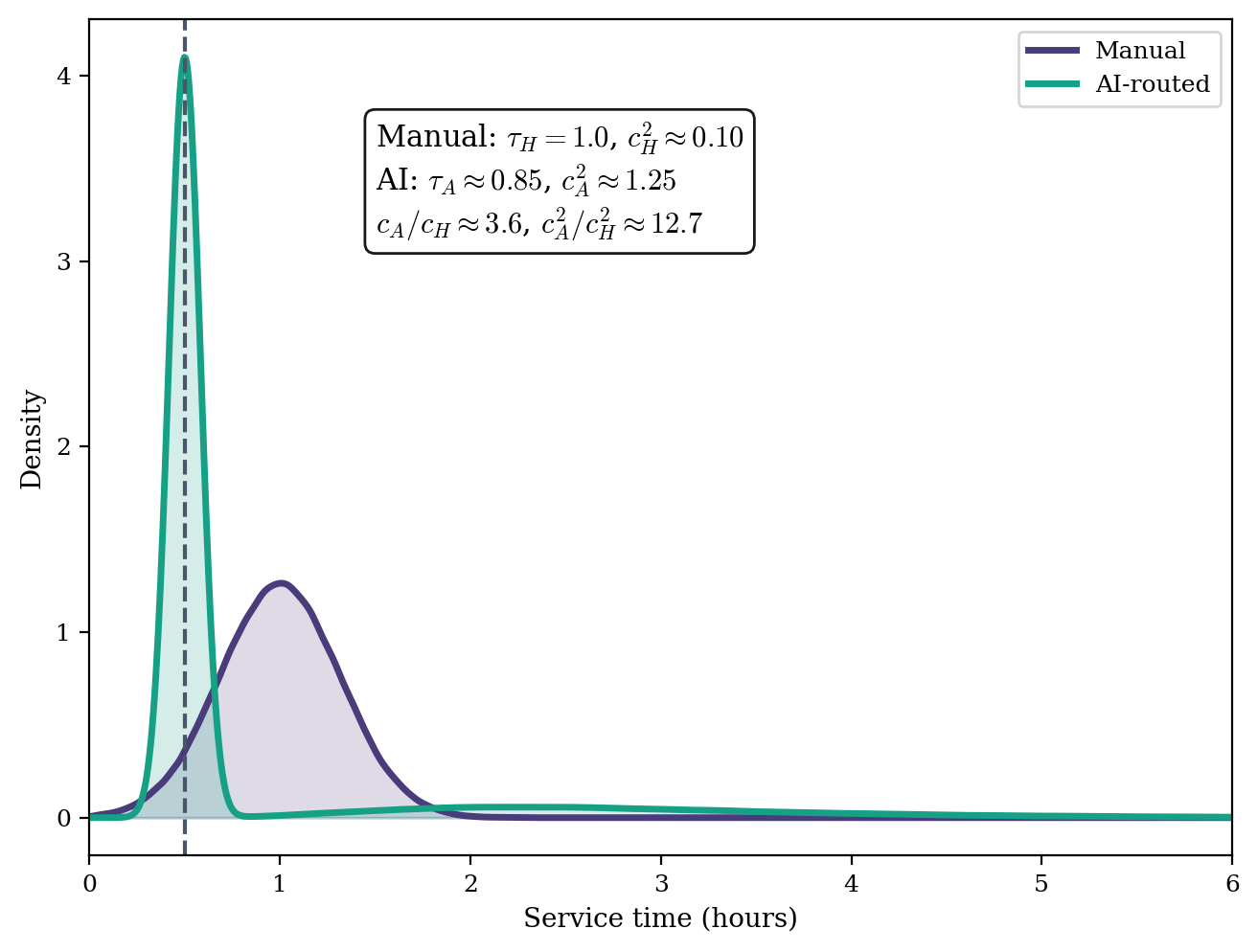}
\caption{Simulated service-time distributions for manual and AI-routed handling.
Manual service time is drawn from a positive, approximately symmetric distribution
with mean \(\tau_H=1\) hour and squared CV \(c_H^2\simeq0.10\).
AI-routed service time is \(T_A=r+MR\), with fixed review time \(r=0.5\), escaped-error
indicator \(M\sim\mathrm{Bernoulli}(0.15)\), and Gamma-distributed rework \(R\).
The AI route has lower mean service time, \(\tau_A\simeq0.85\), but much larger dispersion,
\(c_A^2\simeq1.25\). The sharp peak at \(r=0.5\) corresponds to AI outputs that pass
review without later correction; the long right tail corresponds to escaped errors that
return as rework. 
}
\label{fig:dist}
\end{figure}
The second subtler effect is congestion captured by $(1-\lambda \tau_A/C)/(1-\lambda\tau_H/C)$. This term measures how much farther AI moves the system away from overload. If the manual system is lightly loaded, the denominator \(1-\lambda\tau_H/C\) is not small, so the gain is modest. If the manual system is close to saturation, then \(1-\lambda\tau_H/C\) is close to zero. In that case, even a modest reduction in average service time can create a large operational gain, because it pulls the system back from the edge of instability.

Thus, when the manual system is nearly overloaded, AI can create a larger rework tail and still improve waiting times. When the manual system has plenty of slack, the mean-time saving is less valuable, so even moderate extra variability can produce the variance wedge: AI can reduce the average amount of human time spent per task, \(\tau_A<\tau_H\), and still increase waiting time because the work becomes less predictable. The queue is not slowed down by the typical AI-assisted task; it is slowed down by the occasional task that escapes review, returns later, and consumes a large amount of extra human attention.

Figure~\ref{fig:dist} shows simulated manual and AI-routed service-time
distributions for a representative calibration. The manual distribution is
roughly symmetric around \(\tau_H=1\) hour with \(c_H^2=0.10\). The AI distribution
has a spike-and-tail shape: a sharp peak at the review time \(r=0.5\) hours,
corresponding to cases with no escaped error, plus a long thin tail from rework.
Its mean is lower, \(\tau_A=0.85\) hours, but its squared CV is
much larger, \(c_A^2=1.25\). Thus the AI CV is roughly
3.5 times the manual one, and its squared CV is more than
twelve times larger: rare long rework cases can have
a disproportionate effect on waiting times.

\begin{figure}[htb!]
\centering
\includegraphics[width=0.7\textwidth]{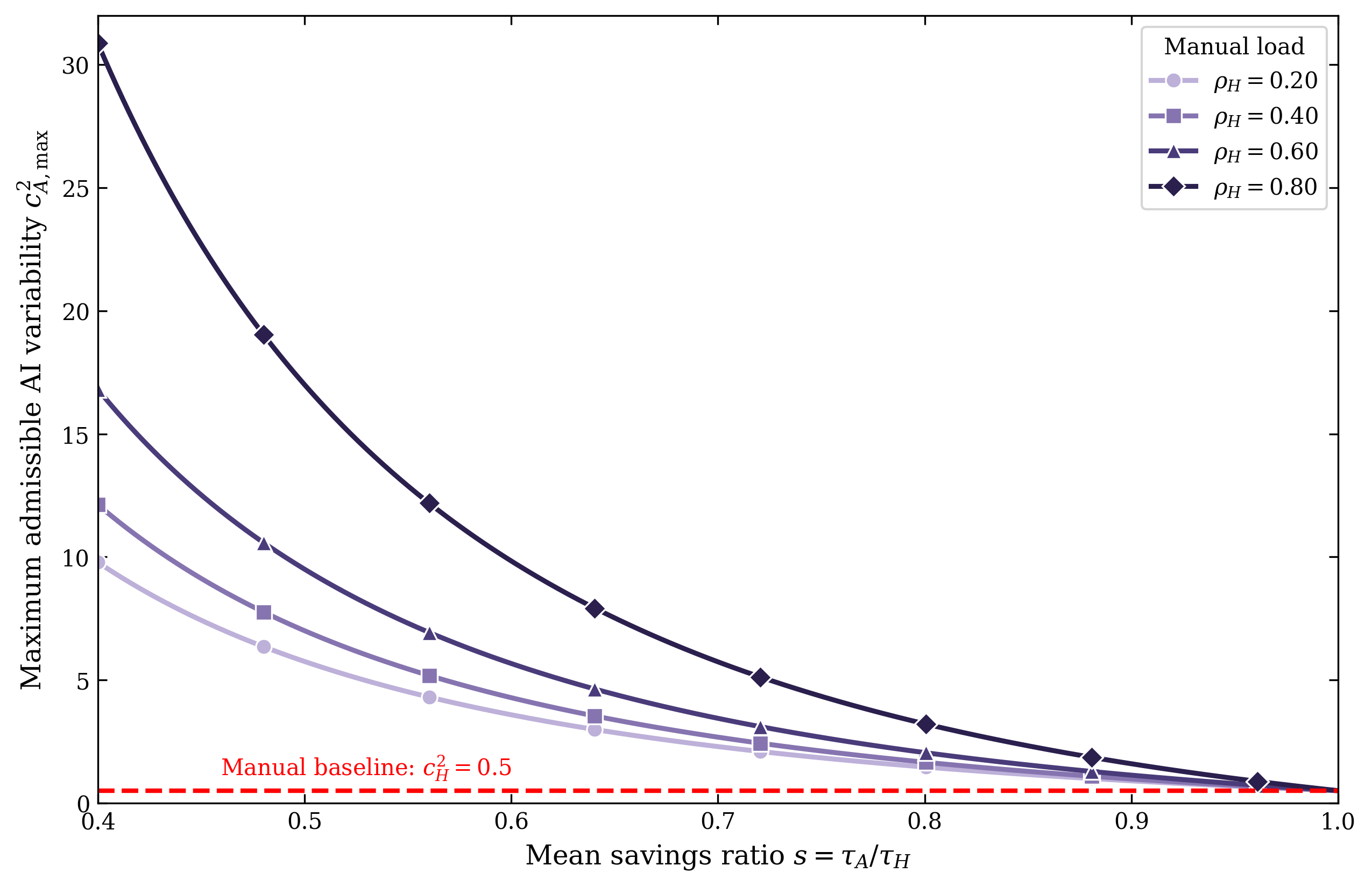}
\caption{Maximum admissible AI variability as a function of the mean-savings ratio
\(s=\tau_A/\tau_H\), computed from equation~\eqref{eq:design}, for four levels of manual-system utilization,
\(\rho_H\in\{0.20,0.40,0.60,0.80\}\), fixing the manual squared coefficient of
variation at \(c_H^2=0.5\). Each curve gives the largest AI squared coefficient
of variation that is still compatible with lower mean waiting time. Points below
a curve correspond to parameter combinations for which AI improves queueing
performance; points above it correspond to the variance wedge, in which AI saves
time on average but increases waiting time. The admissible variance rises sharply
as \(s\) falls, because large mean savings create room for the queue to absorb
substantial rework variability. By contrast, as \(s\to1\), all curves collapse
toward the manual baseline \(c_H^2=0.5\): when AI delivers little mean saving,
its variability can exceed manual variability only marginally before waiting
times worsen. Higher manual utilization shifts the curves upward, showing that
when the manual system is closer to saturation, a given reduction in mean
service time is more valuable and can offset a larger increase in variability.}
\label{fig:design}
\end{figure}

Keeping $c_H^2$ fixed and parametrizing by the mean-savings ratio $s \coloneqq \tau_A/\tau_H$ and the manual-system utilization $\rho_H = \lambda \tau_H / C$, the maximum admissible AI variability for AI to improve waiting time is, from \eqref{eq:trap_CV},
\begin{equation}
\label{eq:design}
{c^2_{A,\mathrm{max}}} \;=\; \frac{1}{s^2} \cdot \frac{1 - \rho_H ~s}{1 - \rho_H} \cdot (1 + c_H^2) \;-\; 1 
\end{equation}
(see Section \ref{app:variancetrap} for a derivation). Figure~\ref{fig:design} plots ${c^2_{A,\mathrm{max}}}$ in Eq.~\eqref{eq:design} as a
function of the mean-savings ratio \(s=\tau_A/\tau_H\), for four illustrative levels
of manual-system utilization, \(\rho_H\in\{0.20,0.40,0.60,0.80\}\), fixing
\(c_H^2=0.5\). Each curve shows the largest AI variability,
\(c_{A,\max}^2\), that is still compatible with lower mean waiting time.

When \(s\) is small, hence AI generates large mean savings,
the variance budget rises sharply: the queue can tolerate substantial AI
variability and still benefit from AI adoption. When \(s\) is close to one, hence AI
saves little average time, ${c^2_{A,\mathrm{max}}}$ collapses toward the manual baseline
\(c_H^2\). In that region, AI variability can exceed manual variability only
slightly before the variance wedge appears. A higher load shifts the entire
schedule upward. Intuitively, when the manual system is already closer to
saturation, a reduction in mean service time is operationally more valuable, so
the queue can absorb a larger increase in AI-induced variability.

AI should, therefore, not be evaluated only by asking
whether it reduces the average time spent on a task. In a queue-bound workflow,
the relevant question is rather whether it improves the service-time distribution in the way the queue needs. Before deployment, an organization should therefore measure three objects: the mean human-attention time under AI, $\tau_A$ ; the variability of
that time, as summarized by the empirical squared CV
\(c_A^2\); and the current load of the manual system, \(\rho_H=\lambda\tau_H/C\).
These quantities are enough to check whether the inequality in ~\eqref{eq:trap_CV} holds.

This diagnostic is simple, but nevertheless quite powerful as it does not require a full structural model of AI errors. It essentially requires logging how long AI-routed cases actually occupy
the human team, accounting for downstream tasks of review when errors escape.

\subsubsection{Partial adoption}\label{sec:partial}

The comparison above contrasts the two extreme cases, no AI routing (\(x=0\)) and full AI routing (\(x=1\)).
Because the model allows partial task routing, it is natural to ask whether an
intermediate adoption level $x^*$ could minimize the overall waiting time, still under the assumption of stable queues. In the baseline homogeneous model, the answer is negative: for a fixed review policy \(r\), the waiting-time objective is monotone in \(x\). The waiting-time optimum is, therefore, ``bang-bang'':
\begin{equation}
\label{eq:bang_bang_x}
x^\star
=
\begin{cases}
1\ ,
&
C(q_A-q_H)+\lambda(q_H\tau_A-q_A\tau_H)<0\ ,\\[4pt]
0\ ,
&
C(q_A-q_H)+\lambda(q_H\tau_A-q_A\tau_H)>0\ ,\\[4pt]
\text{any }x\in[0,1]\ ,
&
C(q_A-q_H)+\lambda(q_H\tau_A-q_A\tau_H)=0 \ ,
\end{cases}
\end{equation}
(see Section \ref{sec:bangbang} for a proof).

This result explains why the pure manual versus pure AI comparison performed in Section \ref{sec:varwedge} is not merely
a modelling convenience. In the homogeneous baseline model, if increasing AI
routing improves (worsens) waiting time, it improves (worsens) it at every adoption level. Keeping the other parameters fixed, Eq. \eqref{eq:bang_bang_x} identifies a possible transition e.g. in the arrival rate $\lambda$, with a critical value
\begin{equation}
    \lambda^\star:=\frac{C(q_H-q_A)}{q_H\tau_A-q_A\tau_H}\ ,\label{eq:lambdastar}
    \end{equation}
between a regime where full AI-handling is favored, and one where full manual handling is favored.

\begin{figure}[htb!]
\centering
\includegraphics[width=0.7\textwidth]{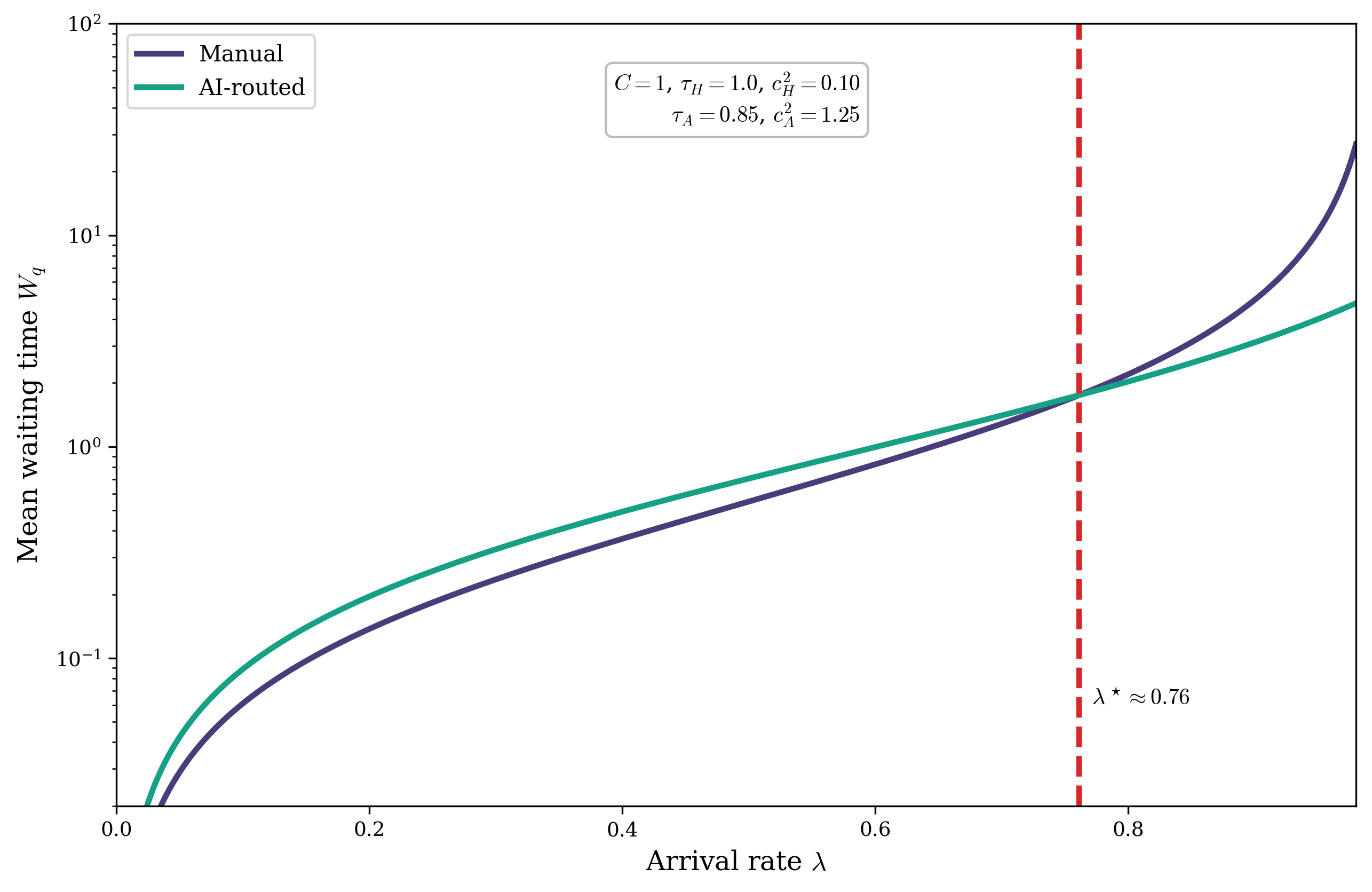}
\caption{Mean waiting time under manual ($W_H$) and AI-routed ($W_A$) handling as the arrival rate
\(\lambda\) varies. The curves are computed from the \(M/G/1\) waiting-time
formula in Eq. \eqref{eq:Wq_model}. Capacity is normalised to \(C=1\). The manual route has mean \(\tau_H=1\) and squared CV \(c_H^2=0.10\), so its second
moment is \(q_H=\tau_H^2(1+c_H^2)\). The AI-routed route has lower mean service
time, \(\tau_A=0.85\), but much higher variability, \(c_A^2=1.25\), giving
\(q_A=\tau_A^2(1+c_A^2)\). At low and moderate arrival rates, the AI curve lies above
the manual curve: although AI saves time on the average task, the larger second
moment created by occasional rework increases waiting time. The curves cross at
\(\lambda^\star\simeq0.76\) as predicted by Eq. \eqref{eq:lambdastar}. Above this point, the manual system is close enough
to saturation that the reduction in mean service time dominates the variance
penalty, and AI-routed handling yields shorter waits. The figure illustrates the
variance wedge: lower average task time need not imply shorter queues.}
\label{fig:trap}
\end{figure}

Figure~\ref{fig:trap} shows the mean waiting time under manual ($W_H$) and AI-routed ($W_A$) handling as the arrival rate of tasks
\(\lambda\) varies. We compute $W_H$ and $W_A$ specializing the \(M/G/1\) formula in equation~\eqref{eq:Wq_model} to the $x=0$ and $x=1$ cases, respectively, varying the arrival rate \(\lambda\), while holding the other parameters fixed. At low and moderate arrival rates, the AI-routed system has longer waits than the manual system. The reason is that the queue is not yet close to capacity, so the mean time saving from AI is not valuable enough to offset the larger second moment created by occasional rework. The two curves cross at $\lambda^\star \simeq 0.76$ as predicted by Eq. \eqref{eq:lambdastar}, which marks a regime transition. Below this threshold, AI reduces average human time per task, but increases average waiting time. Above it, the manual system is close enough to saturation that the lower mean service time of the AI route dominates its higher variability. Thus, AI becomes beneficial only when the manual queue is already under substantial pressure.

A critical intermediate routing fraction $x_c$ will be determined in Sec. \ref{sec:queuestabilizing2} that answers a substantially different question, though: how much load should one divert from manual to AI handling in order to turn an \emph{unstable manual queue} into a stable one? Intermediate routing
fractions may also become important once tasks differ in their AI suitability, review
capacity is constrained by task type, or if AI adoption affects future review skills. These extensions will not be considered here.

We now turn to the discussion of two direct implications of our main results.

\subsection{Implication $\# 1$: review threshold under congestion}\label{sec:thresholdcong}
Let us now focus on the fully AI-routed workflow $(x=1)$.
So far, we have treated the review effort $r$ as fixed. In practice, reviewers do not spend the same amount of time on every AI output: they use warning signs in the task and in the draft itself to strategically decide which outputs may deserve closer scrutiny. We therefore replace the fixed review time with a selective checking rule that depends on an observable risk signal. 

After a cursory look at the AI draft, an experienced reviewer can often gauge its level of risk.
Some clues come from the task itself: a routine request is usually safer than a
novel or high-stakes one. Other clues come from the draft: vague citations,
unusual confidence, missing calculations, or inconsistencies with the source
material may all suggest that the output deserves closer scrutiny.

We collect these clues in a signal \(s\). The signal is not assumed to be perfectly correlated with the actual danger:
it only changes the reviewer's assessment of risk. Let \(\pi(s)\in[0,1]\) denote
the reviewer's posterior probability that the AI draft contains an error after
observing the signal \(s\). Across tasks, signals are distributed according to a density \(\mu(s)\) defined over a support $\bar S$.
If an error escapes review, it creates an expected loss \(K>0\), which may include
correction time, escalation, customer harm, legal exposure, or audit costs. We call \(\theta>0\) the congestion cost of using reviewer time. It captures the fact that time spent checking one AI output occupies the same human bottleneck needed to process the rest of the queue: every additional minute of review delays the tasks waiting behind it. Thus \(\theta\) is low when reviewers have spare capacity, but rises when the queue is congested. We initially treat $\theta$ as a parameter fixed at the outset, and then return to its endogenous determination later on.

For a draft with posterior error risk \(\pi(s)\), review has two opposing effects.
It consumes human resources that are often scarce, but it lowers the probability that an error
survives. We assume that each unit of review reduces the remaining error risk at
rate \(\kappa\), so that early review catches the easiest errors, while later review
still helps but by progressively smaller amounts.

The resulting policy is selective verification: the reviewer checks the draft only
if the perceived error risk is high enough. Our model predicts a review threshold (see Section \ref{sec:reviewthreshold} for details)
\begin{equation}
    \pi^\star(\theta)
=
\frac{\theta}{\kappa K}\ .\label{eq:pistar}
\end{equation}

Drafts whose posterior risk $\pi(s)$ is below $\pi^\star$ are passed through without scrutiny; drafts above the threshold
receive review effort that grows logarithmically with their perceived risk, according to the formula
\begin{equation}
\label{eq:r_starMain}
r^\star(s; \theta)
=
\frac{1}{\kappa} \max\!\left\{0,\; \log\frac{\kappa K \pi(s)}{\theta}\right\}\ .
\end{equation}

Eq. \eqref{eq:pistar} shows that the review threshold rises with the congestion cost of reviewer time, \(\theta\), and falls with both the cost of an escaped error, \(K\), and the effectiveness of review, \(\kappa\). Thus reviewers become more selective when attention is scarce, but scrutinize more drafts when errors are more costly or review is more effective. Formula \eqref{eq:r_starMain} shows that review effort increases with the perceived risk \(\pi(s)\) and with the cost of an escaped error \(K\), because both make checking more valuable, but decreases as the congestion cost of reviewer time \(\theta\) rises; the parameter \(\kappa\) both scales the effectiveness of review and appears in the logarithm, so higher review effectiveness generally means that less time is needed to achieve a given reduction in risk. This connects
the model to the inspection literature, where a central question is how to allocate
limited checking effort across many items when some are more likely to be defective
than others \cite{lindsay1964,tapiero1996}.

The formulas above describe how a reviewer behaves \emph{for a given congestion cost \(\theta\) of reviewer time}. The remaining step is to determine where this cost comes from. In the queueing system, reviewer time is costly not only because it consumes effort, but because it negatively affects the ability of the same team to process the other tasks waiting in line. We therefore link \(\theta\) to congestion: an additional unit of review increases the average amount of human attention required per task, which in turn increases waiting time for the rest of the queue. The more heavily loaded the team is, the larger this delay effect becomes. In Section~\ref{sec:reviewthreshold} we formalize this idea by setting \(\theta\equiv\theta^\star\) as the solution of a fixed-point equation $\theta^\star=\Phi(\theta^\star)$ (see Eq. \eqref{eq:verification_fixed_point}), 
where \(\Phi(\theta)\) is the marginal queueing cost generated by an additional unit of mean human-attention time per task: the derivative of waiting time with respect to the mean service requirement, multiplied by the arrival rate $\lambda$ and by the waiting-cost parameter \(c_w\), which measures how costly it is for a task to spend one additional unit of time waiting in the queue.

This closes the feedback loop between review and congestion. The review rule determines how much time is spent checking AI outputs; that review time contributes to the load on the human bottleneck; and the resulting congestion determines how costly further review time becomes. When congestion raises \(\theta^\star\), the threshold $\pi^\star(\theta^\star)$ in Eq. \eqref{eq:pistar} moves upward. More drafts therefore fall below the threshold and receive no review. Even for drafts that remain above the threshold, the rule Eq. \eqref{eq:r_starMain} evaluated for $\theta=\theta^\star$ gives lower review effort at any fixed perceived risk \(\pi(s)\). Thus congestion does not merely delay work mechanically: it changes the verification policy itself, making reviewers more selective about which AI outputs they check.

This provides a queueing-based interpretation of under-verification. It is consistent with Ref. \cite{parasuraman2010}'s observation that automation-induced complacency intensifies under load, and with Ref. \cite{bainbridge1983}'s argument that automation reorganizes rather than removes human workload. Here, however, the effect is not introduced as a cognitive bias: it arises because careful checking consumes scarce attention in a congested system.

The mechanism becomes especially consequential in workflows where risk and congestion tend to increase together. This co-movement is not assumed by the threshold formula itself, but it is common in practice: peak-hour customer service, end-of-quarter reporting, urgent legal review, and clusters of complex cases are often both busier and harder than ordinary periods. In such settings, the distribution of signals \(\mu(s)\) shifts toward riskier drafts at the same time as the arrival rate \(\lambda\) raises the congestion cost \(\theta^\star\). The model then predicts a troubling response: the review threshold rises precisely when more outputs would deserve closer scrutiny. Reviewers facing a congested queue and scarce time become more selective on which outputs to check at the very moment when careful checking would be needed the most. 

\begin{figure}[htb!]
\centering
\includegraphics[width=\textwidth]{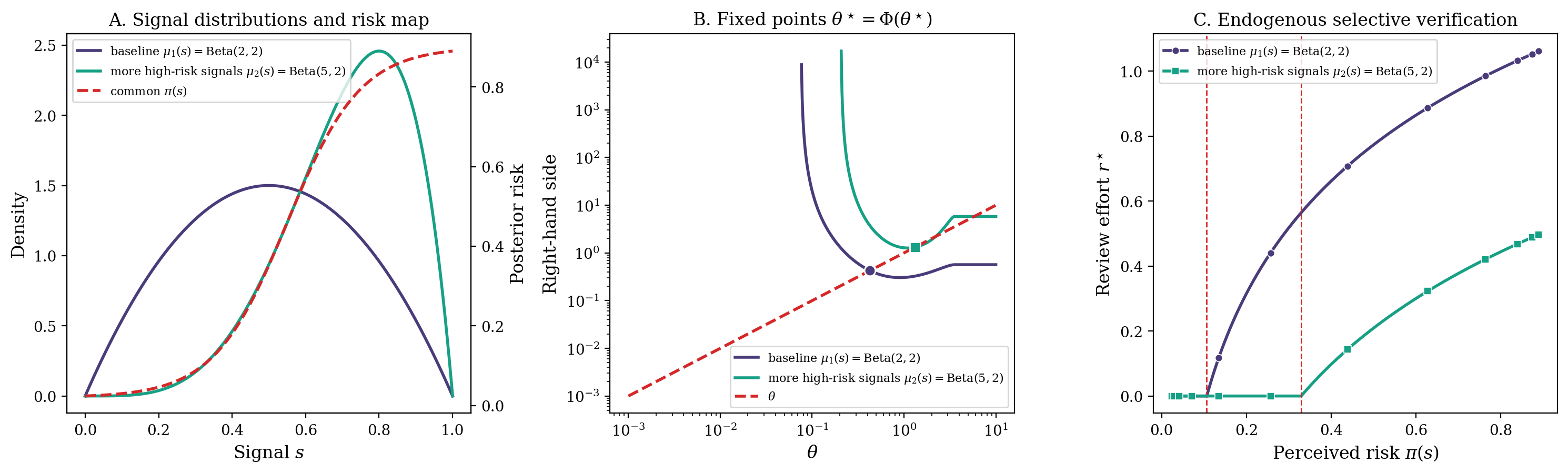}
\caption{Congestion-driven selective verification under two signal distributions. 
The posterior error-risk map is held fixed at 
$\pi(s)=0.02+0.88/[1+\exp(-10(s-0.55))]$, while the signal density is either 
$\mu_1(s)=\mathrm{Beta}(2,2)$ or $\mu_2(s)=\mathrm{Beta}(5,2)$; the latter shifts mass toward larger $s$, making high-risk drafts more frequent. 
Common parameters are $C=1.00$, $\lambda=0.75$, $c_w=0.50$, $\kappa=2.00$, $K=2.00$, $\mathbb{E}[R]=1.50$, and $\mathbb{E}[R^2]=4.00$. 
Panel A shows the two signal densities and the common risk map. 
Panel B solves $\theta^\star=\Phi(\theta^\star)$ (see Eq. \eqref{eq:verification_fixed_point}), where 
$\Phi(\theta)=c_w\lambda\,\partial W_q/\partial \tau_A$ and 
$W_q=\frac{\lambda q_A(\theta)}
{2C(C-\lambda \tau_A(\theta))}$ (see Eqs. \eqref{eq:Wq_model}, \eqref{eq:routemoment1} and \eqref{eq:routemoment2}). In case of more than one intersection, we pick the smallest root.  
Panel C plots the induced review rule 
$r^\star(s;\theta)=\kappa^{-1}\max\{0,\log(\kappa K\pi(s)/\theta)\}$, with residual escaped-error probability 
$\pi(s)\exp[-\kappa r^\star(s;\theta)]$. 
For $\mu_1$, $\theta^\star=0.426$, $\pi^\star=0.106$, $\tau_A=0.671$, $q_A=0.995$, $\rho=0.503$, and $W_q=0.751$. 
For $\mu_2$, $\theta^\star=1.316$, $\pi^\star=0.329$, $\tau_A=0.831$, $q_A=1.770$, $\rho=0.623$, and $W_q=1.762$. 
When high-risk drafts become more frequent, the fixed point shifts to a higher price of attention and a higher review threshold.}
\label{fig:theta_fixed_point}
\end{figure}

Figure~\ref{fig:theta_fixed_point} illustrates the verification equilibrium implied by Eq.~\eqref{eq:verification_fixed_point} for two signal environments. In both cases, the posterior risk map $\pi(s)$ is held fixed and increasing in the signal $s$, while the distribution of signals changes: the baseline density $\mu_1(s)=\mathrm{Beta}(2,2)$ places most mass on intermediate signals, whereas $\mu_2(s)=\mathrm{Beta}(5,2)$ shifts mass toward larger signals and therefore makes high-risk drafts more frequent. Panel A shows that the two workflows share the same mapping from signal to perceived error risk but differ in how often risky signals occur. Panel B displays the corresponding fixed-point problem for the cost $\theta$ of reviewer attention. In the calibration shown, moving from $\mu_1$ to $\mu_2$ raises the equilibrium verification cost from $\theta^\star\simeq 0.426$ to $\theta^\star\simeq 1.316$, and the associated review threshold from $\pi^\star\simeq 0.106$ to $\pi^\star\simeq 0.329$. Panel C translates these fixed points into review policies: when high-risk drafts become more common, congestion raises the cost of attention, so the reviewer becomes more selective, passing through more intermediate-risk drafts without scrutiny and exerting less review effort at a given perceived risk. The figure therefore shows the central selective-verification mechanism: congestion endogenously raises the cost of review time, shifting the review threshold upward precisely in environments where risky drafts are more frequent.
\begin{figure}[htb!]
\centering
\includegraphics[width=0.7\textwidth]{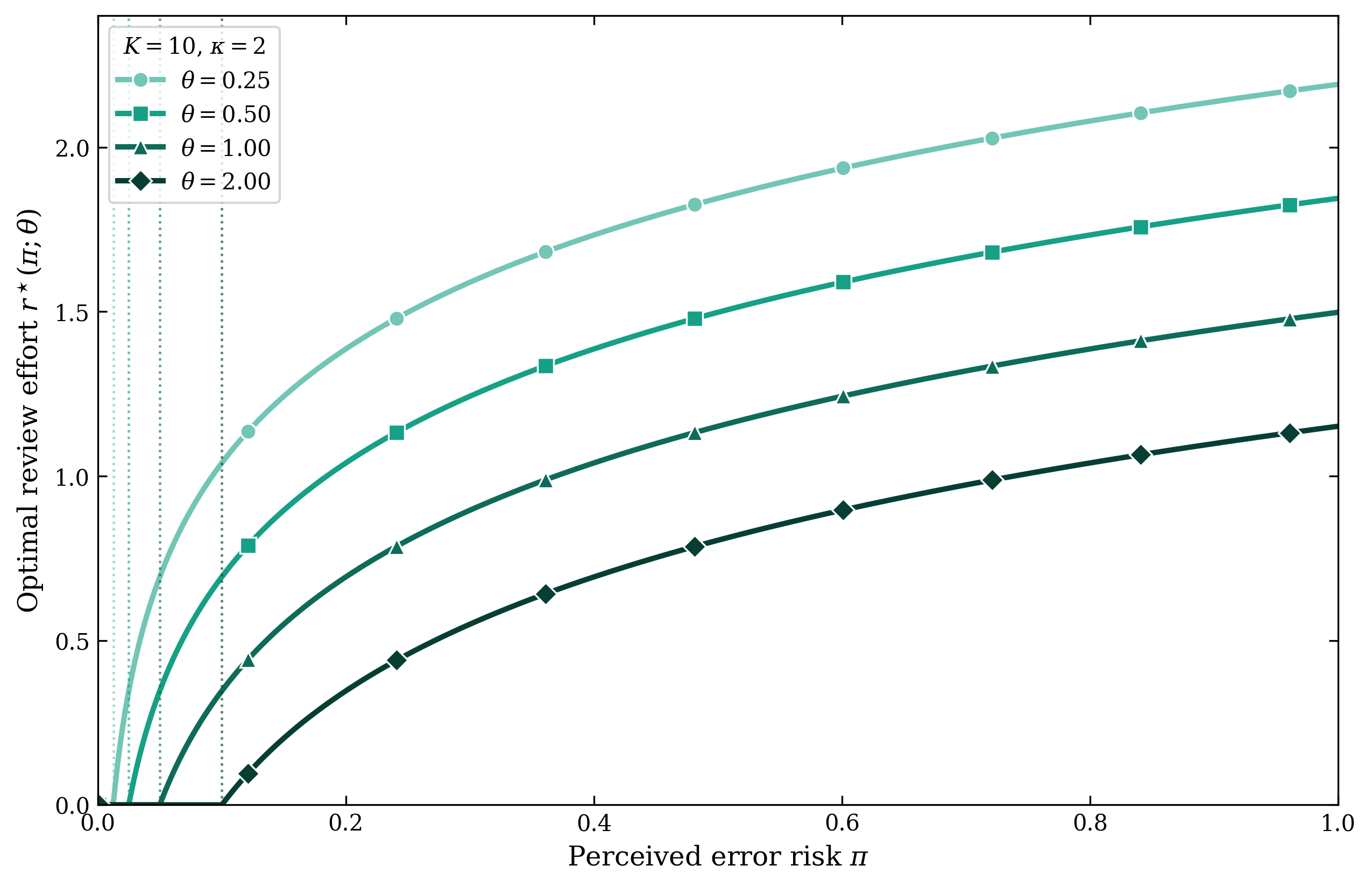}
\caption{
Optimal review effort as a function of the perceived error risk \(\pi\).
The curves plot the optimal review rule derived in the Material and Methods for
four values of the cost of reviewer time, \(\theta\), fixing \(K=10\) and
\(\kappa=2\). The dotted vertical lines mark the corresponding review thresholds.
For risks below the threshold, the optimal review effort is zero. Once the
perceived error risk exceeds the threshold, review begins and increases
logarithmically with \(\pi\). Higher values of \(\theta\) make reviewer time
more expensive, shift the threshold to the right, and reduce review effort at
every given level of perceived risk.
}
\label{fig:verification}
\end{figure}
Figure~\ref{fig:verification} plots the optimal review effort as a function of
the perceived error risk \(\pi\in[0,1]\). This risk is induced by the signal
\(s\): after observing \(s\), the reviewer assigns probability
\(\pi:=\pi(s)\) to the event that the AI draft contains an error. Higher $\theta$ corresponds to a more congested system, in which any minute spent reviewing a draft delays every other task in the queue. The threshold moves rightward, so a larger range of drafts, now including some with non-trivial error risk, is passed through unchecked. For drafts that remain above the (now higher) threshold, review effort is also reduced at every level of perceived risk: each curve sits below the curves corresponding to lower $\theta$ values.

\subsection{Implication $\# 2$: queue-stabilizing effect of AI}
\label{sec:queuestabilizing2}

The variance wedge effect concerns the comparison of a fully manual vs. a fully AI-assisted workflow, when both systems are assumed to be in their stable regime $(\rho_H<1, \rho_A <1)$. A different question is whether AI can help stabilize an overloaded workflow. Suppose that the fully manual system is overloaded (i.e., $\lambda \tau_H > C$)
so that the average incoming human-attention demand exceeds the capacity of the workflow. In this case, the manual queue does not have a finite steady-state waiting time. A central result of our model is that AI can stabilize the workflow only if routing enough tasks through the AI workflow reduces average human-attention demand below capacity. Recalling that $m(x;r)$ is the average amount of human-attention time required by each incoming task, the stability condition of the mixed manual-AI queue \(\lambda m(x;r)<C\) gives from Eq. \eqref{eq:m} below
\begin{equation}
\label{eq:xc}
x>x_c(r)
\coloneqq
\frac{\lambda \tau_H-C}
{\lambda(\tau_H-\tau_A)}\ .
\end{equation}
Thus \(x_c(r)\) is the minimum fraction of tasks that must be routed via AI to make the queue stable. It is positive when the manual system is overloaded, but it also requires that AI uses less human attention on average than manual handling, \(\tau_A<\tau_H\). If even the fully AI-routed system is overloaded, \(\lambda \tau_A\geq C\), then no adoption share can stabilize the queue.

Spelling out \(\tau_A\) using Eq. \eqref{eq:ar1}, the basic condition for ``helpful'' AI assistance is
\begin{equation}
\label{eq:rescue}
\tau_A=r+p(r)\mu_R<\tau_H\ .
\end{equation}
This condition is substantially more stringent than simply observing that AI can generate a draft quickly. The relevant quantity is rather the \emph{total} human attention consumed by the AI route, which includes the time spent reviewing the draft plus the expected rework created by errors that escape review. An AI system may produce a first draft in seconds and still fail Eq.~\eqref{eq:rescue} if the resulting review and rework burden exceeds the time needed to do the task manually.

This stabilization result should be distinguished from the variance wedge condition \eqref{eq:trap_CV}. Stability is a mean-load condition: on average, does the workflow send less human-attention demand to the workflow than it can process? Waiting-time performance is a stricter condition. Among stable systems, the mean waiting time also depends on the second moment of service requirements. Hence AI can reduce mean load enough to stabilize an otherwise overloaded workflow, while still producing longer waits than a manual or lower-variance alternative would produce in a stable comparison. In short, the queue may be rescued on the first moment and damaged on the second.

The ``bang--bang'' adoption result in Sec. \ref{sec:partial} and the stabilization threshold derived here refer to different scenarios, and the apparent clash between the two is readily resolved. The bang--bang result is a statement about waiting-time optimization \emph{within the stable region}. For fixed review effort \(r\), the waiting-time expression \(W_q(x;r)\) is monotone in the AI-adoption fraction \(x\) wherever the queue is stable. Hence, an intermediate value of \(x\) cannot minimize the waiting time, except in the knife-edge case where the derivative vanishes: the waiting-time optimum lies at a boundary of the stable feasible set. The critical fraction \(x_c(r)\), by contrast, is not an optimum but a stability threshold. It becomes relevant when the  fully manual system is overloaded, \(\lambda\tau_H>C\), so that \(x=0\) is not a stable feasible point at all. In that case, the stable region begins only once AI adoption is large enough to reduce the average human-attention demand below capacity, namely for \(x>x_c(r)\). Thus, \(x_c(r)\) marks the lower boundary of the stable region, not the adoption level that minimizes the waiting time. Once \(x\) lies above this boundary, the same monotonicity logic behind the bang--bang result applies on the restricted stable interval \(x\in(x_c(r),1]\). If the waiting time decreases with \(x\), the best stable choice is full AI routing, \(x=1\). If the waiting time increases with \(x\), then additional AI routing worsens waiting time once stability has been restored; the infimum would be approached as \(x\downarrow x_c(r)\), although waiting time itself diverges at the stability boundary, so in practice one would choose an adoption share slightly above \(x_c(r)\) that provides a safety margin. The two results are therefore complementary: \(x_c(r)\) says how much AI adoption is needed to prevent the queue from exploding, while the bang--bang condition says in which direction waiting time moves once the system is stable.

The operational implication is that the system's boundary matters. The condition in Eq. \eqref{eq:rescue} is only as reliable as the measurement of the rework behind it. If escaped errors are pushed outside the measured workflow---onto customers, downstream teams, auditors, or future reporting periods---then the AI route will appear artificially cheap, because part of the true human-attention burden is not captured by \(\tau_A(r)\). The stabilization condition may then appear to hold even though the organization has simply moved work elsewhere. If escaped errors return to the same team, by contrast, the rework burden \(\mu_R\) enters \(\tau_A(r)\) directly, and AI is stabilizing only when the sum of review and expected rework genuinely consumes less human attention than manual handling. To know whether AI has increased capacity rather than displaced effort, organizations must, therefore, also measure the extra work created when AI outputs fail.

\section{Discussion and Conclusion}\label{sec:conclusion}

The recent empirical literature on generative AI deployment in the workplace often documents substantial per-task productivity gains, alongside  heterogeneity across workers, tasks, and domains. The argument of this paper is that even where average gains are real, they may not translate into improvements in queuing-theoretic measures of system responsiveness, and may sometimes produce the opposite effect. While AI shortens the typical task completion by drafting it cheaply, it introduces a tail of long tasks via the rework mechanism triggered by escaped errors. Since waiting time in a queue scales with the second moment of service time, this reshaping can simultaneously raise productivity (tasks per hour) and degrade responsiveness (waiting time per task in queue).

We have made this argument quantitative through an \(M/G/1\) queue model of AI-assisted work with rework and selective verification. Three results follow from the same underlying mechanism: AI changes the full service-time distribution faced by scarce human attention, not only its mean. First, the variance wedge condition shows that AI can reduce the average human time per task, while increasing the queueing delay, because rare rework cases raise the second moment of service time. Second, the selective-verification rule shows that congestion changes review behavior itself: when reviewer time becomes more costly, the risk threshold for checking AI outputs rises and review effort falls even for drafts that are still inspected. Third, the stabilization condition shows that AI can rescue an overloaded workflow only if adoption is high enough and the AI route genuinely reduces total human-attention demand, including review and expected rework. Together, these results show why mean handle time is an incomplete and often misleading measure of AI productivity in queue-bound workflows: the system-level outcome depends on load, variability, and the extra work created when errors return.

Operationally, the model reduces the deployment test for AI assistance in an organization to measuring the realized distribution of human-attention time per task, including any rework, and the arrival rate of tasks. If the condition in Eq.\eqref{eq:trap_CV} fails, AI will effectively lengthen queues even if the average task time improves.

The broader lesson is that task-level productivity and workflow responsiveness are different metrics. They diverge most sharply when AI delivers modest mean savings, but creates rare costly failures in systems already close to capacity. In such settings, a faster typical task completion can coexist with a slower overall workflow. Moreover, our work argues that classical queueing theory is the right language to frame questions about stability and efficiency of AI-assisted workflows.

The model deliberately works at a level of abstraction that omits several features of real organizations. Some omissions are clearly costly, others less so. We outline future research directions worth exploring.

\paragraph{Layered human review.} 
We represented the human side of the workflow as a single team characterized by fixed human-attention resources with capacity \(C\). Real organizations are more structured. Routine cases may be handled by junior staff, while difficult cases, escalations, or approvals may require senior reviewers. AI adoption can therefore shift pressure across layers of the organization: it may reduce the amount of junior production work while increasing the demand for senior verification as AI can substitute for routine drafting more easily than for expert judgment. A two-tier extension of our model, with separate ``junior'' and ``senior'' review stages, would capture this richer dynamics. AI could relieve one layer while overloading another. We expect the qualitative mechanisms developed here (the variance wedge, threshold under-verification, and the role of rework) to carry over, but the redistribution of load across skill tiers would become an additional object of analysis. We view this as one of the most important future extensions.

\paragraph{Bursty arrivals.} Most of our explicit calculations use the M/G/1 Pollaczek--Khinchine formula, which assumes Poisson arrivals. Real workflows are bursty, with $c_a^2 > 1$. From the Kingman approximation in Eq. \eqref{eq:Kingman}, burstiness worsens the waiting-time consequences of both mean load and service-time variability, though it does not change the first-order stability condition $\rho < 1$. The variance wedge is expected to be more severe under burstiness, and the equilibrium cost $\theta^\star$ to be higher. The development of the full theory and implications in the case of irregular arrivals is also an interesting topic for future work.

\paragraph{Adversarial errors.}
So far we have treated the error process as fixed: review effort lowers the probability that an error escapes, but the errors themselves are not strategically placed. This is a reasonable first approximation for many internal workflows, but not for settings in which an external party seeks to actively probe the verification system and its vulnerabilities. In security review, content moderation, fraud detection, or litigation, failures are not drawn passively from a stable distribution; they are selected to exploit predictable gaps in scrutiny and to gain an advantage. In such environments, a deterministic threshold rule can turn into a liability: once adversaries learn which cases receive little review, they can concentrate attacks just below the threshold. A natural extension is therefore to replace the deterministic checking rule with a randomized or mixed-inspection policy, so that verification remains partly unpredictable while still allocating more attention to higher-risk cases. This would connect the present model to adversarial inspection games  \cite{dresher1962sampling,avenhaus2002inspection} and to modern security games in which limited defensive resources are randomized to prevent strategic exploitation \cite{kiekintveld2009computing}. This re-framing would be essential for applications where the rework tail is generated not only by accidental AI errors, but by strategic attempts to evade or circumvent review.

\paragraph{Severity of escaped errors.}
The baseline model assumes that review effort changes the probability that an
error escapes, but not the severity of the error conditional on escape. Formally,
the rework burden \(R\) is taken to be independent of the review time \(r\). This
keeps the moment formulas transparent: review lowers the term \(p(r)\), while the
conditional rework moments \(\mu_R\) and \(\mu_{R,2}\) are fixed.

This is a crude simplification. In some workflows, more careful review may catch the
most severe errors first, so that errors escaping after longer review are less
costly on average. In that case the relevant conditional moments,
\(\mathbb E[R\mid M=1,r]\) and \(\mathbb E[R^2\mid M=1,r]\), would fall with
\(r\). Allowing for this behavior would weaken the rework tail and make AI less
likely to fall into the variance wedge. It would not remove the mechanism entirely, though: the
queueing condition would still depend on the realized second moment of
AI-routed service time after review. Empirically, the question is therefore not
whether review reduces error probability alone, but whether it also thins the
tail of escaped-error rework.

\paragraph{Endogenous AI quality.}
We have treated AI quality as fixed: the baseline error rate \(p_0\), the irreducible error rate \(p_\infty\), and the rework distribution do not change with deployment. This is a useful static benchmark, but in practice these quantities may evolve. Model improvements, better prompting, domain-specific fine-tuning, retrieval systems, and feedback from reviewers can all lower the probability that an AI output contains an error or reduce the severity of the errors that remain \cite{ji2023}. A natural extension is therefore to make AI quality a state variable. For example, \(p_0\), \(p_\infty\), or the moments of \(R\) could decline over time as the system is corrected, monitored, and adapted to the workflow. This would turn the variance wedge into a dynamic object: an AI deployment that initially fails the queueing test might become beneficial after sufficient learning, while one that appears safe at launch could become risky if rising task volume exposes rare failure modes. The key empirical question would then be whether improvements in AI quality occur quickly enough to reduce the rework tail before congestion and accumulated errors dominate the workflow.

\paragraph{Routing fraction.}
In the baseline model, the adoption share \(x\) is deliberately simple: tasks are homogeneous, the review rule is fixed, and changing \(x\) only mixes the manual and AI-assisted service-time distributions. Under these assumptions the waiting-time objective is monotone on the stable region, so the optimal routing rule is ``bang--bang'' instead of leading to an intermediate value $x^\star\in (0,1)$. This should not be read as saying that intermediate adoption levels are unimportant in practice. Rather, it identifies which assumptions make them disappear. Intermediate routing fractions become meaningful once tasks differ in their suitability for AI assistance. Routine, low-risk, or highly standardized tasks may have low review and rework costs under AI, while novel or high-stakes tasks may remain better handled manually. In that case, an interesting tweak in the model would be a routing policy that assigns tasks to AI or manual handling based on observable features of the case. 

A richer routing model would also allow review capacity to be type-specific. For example, AI may reduce the load created by routine production tasks while increasing demand for senior review, escalation, or audit. In this setting, the optimal fraction of tasks to route via AI will depend not only on whether AI reduces average human time, but on where in the organization the saved and added work occur. Finally, routing may have dynamic effects: using AI on more tasks may generate feedback data and improve future AI quality, but it may also change the amount of practice reviewers receive on manual cases. These extensions would turn \(x\) from a static mixing parameter into an operational policy variable. The simple bang--bang result is therefore best interpreted as a benchmark: with homogeneous tasks and a single pool of human attention, there is no intrinsic value to partial adoption; partial adoption becomes important once heterogeneity, capacity constraints by task type, or learning over time are introduced.

\paragraph{Is the mechanism AI-specific?} As we have already remarked, the queueing mechanism in this paper is not unique to generative AI. Any technology or process that
lowers the typical service time while increasing the probability or severity of
rework can generate a similar variance wedge: outsourced service providers, automated decision rules, low-code templates, junior-worker delegation, robotic process automation, spellcheck/autocomplete, rule-based triage systems, and even poorly designed checklists can all exhibit some or all of these features. We view this as a strength rather than a weakness of the framework:
our result is not a claim about the `ontology' of generative AI, but about the operational
signature of a class of technologies.

What makes generative AI a particularly natural application is the combination
of four features. First, AI often reduces the time required to produce a first
draft. Second, the draft may subtly conceal plausible-looking errors that are costly to detect.
Third, human review is still required, and review competes with other tasks for
scarce expert attention. Fourth, escaped errors often return to the same
team as correction, escalation, customer complaint, legal exposure, or
audit work. These features map directly onto the parameters of the model:
review effort \(r\), residual escape probability \(p(r)\), and rework burden
\(R\).

Thus the model should be read as a queueing theory of AI-assisted work, not as a
claim that only AI can create such queues. The AI-specific contribution is to
show how a now-common form of AI deployment (cheap drafting followed by scarce
human verification) can bear the hallmark service-time paradox that classical queueing
theory warns against.

\paragraph{Demand response.}
We have so far treated the arrival rate \(\lambda\) as fixed, but in many real settings the queue also changes the demand it receives. If customers wait too long, they may abandon the service, switch provider, call again later, or open a second channel; if internal users see a tool as slow or unreliable, they may route work around it. These responses can either drain the queue or feed it further, depending on the setting. A useful extension would therefore let arrivals depend on the waiting time generated by the system itself. This would make the model closer to familiar congestion phenomena: long waits change behavior, and changed behavior alters the load faced by the queue. The main question would no longer be only whether AI reduces service time or increases rework, but whether the combined system settles into a high-throughput, low-delay regime or into a self-reinforcing congested regime in which delays, rerouting, and repeated contacts sustain the load. This is especially relevant for AI deployments that change user behavior, for example by making it easier to submit requests, encouraging more marginal cases to enter the workflow, or causing failed outputs to generate repeat interactions.

\paragraph{Empirical calibration.}
The quantities in the model are meant to be operationally measurable, but not all are equally easy to measure. The manual baseline requires the distribution of human-attention time under existing practice, summarized by \(\tau_H\) and \(c_H^2\). The AI route requires the corresponding distribution after AI deployment, including review time, corrections, escalations, complaints, audits, and any later work caused by outputs that initially passed review. From these logs one can estimate \(\tau_A\), \(c_A^2\), the residual error curve \(p(r)\), the mean rework burden \(\mu_R\), and the review-efficiency parameter \(\kappa\). The most important empirical point is that these objects cannot be inferred from prompt-level speed or benchmark accuracy alone: they require following tasks after handoff and recording whether apparently completed outputs return as work. The hardest quantity to estimate is \(c_A^2\), because it is driven by rare tail events and therefore requires enough observations, and a long enough measurement window, to capture downstream rework. A useful empirical agenda would therefore combine time stamps, reviewer effort, error flags, and rework links at the task level, so that AI deployments can be evaluated on the full service-time distribution rather than on average handle time alone.

\paragraph{Full distribution of waiting time.}
We have focused on the mean waiting time \(W_q\), but the classical \(M/G/1\) theory is not limited to means. The Pollaczek--Khinchine result can be stated at the level of the Laplace--Stieltjes transform of the stationary waiting-time distribution; the familiar mean formula is obtained by differentiating that transform \cite{cohen1969,kleinrock1975queueing,asmussen2003applied}. Equivalently, the waiting-time distribution can be characterized through Pollaczek--Khinchine-type integral equations \cite{neuts1986generalizations}, and numerical methods exist for computing waiting-time distributions and cumulants from Pollaczek formulas \cite{abate2000calculation}. Thus our use of \(W_q\) is a first step towards linking classical queueing theory to the properties of AI-assisted workflows, not the end of the story: moving from mean service time to mean waiting time already brings the rework tail into the analysis, while a fuller treatment could study variance, higher cumulants, or how tail probabilities are affected by rare occurrences and more unusual tasks.

\paragraph{Connecting signal risk and arrival load.}
In the present model, the arrival rate \(\lambda\) and the signal distribution \(\mu(s)\) are treated as separate entities: \(\lambda\) determines how many tasks enter the queue, while \(\mu(s)\) determines how risky those tasks appear once they arrive. In many workflows, however, these two objects are likely to move together. Busy times often bring a different mix of tasks: peak-hour support may involve more angry or complex customers, end-of-quarter reporting may bring more unusual cases, and urgent legal or security review may combine higher volume with higher stakes. A natural extension would therefore replace the fixed signal distribution \(\mu(s)\) with a load-dependent distribution \(\mu(s\mid \lambda)\), or equivalently allow the posterior-risk environment to shift as the arrival rate changes. This would make explicit the link between congestion and the level of human attention required per task. When a rise in \(\lambda\) also shifts mass toward higher-risk signals, the system is hit twice: more tasks arrive, and each task is more likely to require careful review. At the same time, congestion raises the cost of reviewer attention, pushing the review threshold upward. Such a model would better capture the operationally important case in which the periods when scrutiny is most needed coincide with it being most expensive.

\section{Materials and Methods}

\subsection{Model specification and service-time moments}
\label{sec:MM_model_primitives}

The total human-attention cost of an AI-routed task is defined in Eq.~\eqref{eq:T_A},
\begin{equation}
T_A=r+MR\ .
\end{equation}
Taking moments gives
\begin{align}
\tau_A &:=\tau_A(r) = \mathbb E[T_A]=r+p(r)\mu_R\ ,\label{eq:ar1}\\
q_A &:=q_A(r)=\mathbb E[T_A^2]=r^2+2rp(r)\mu_R+p(r)\mu_{R,2}\ .\label{eq:a2}
\end{align}
The AI-route squared CV is therefore
\begin{equation}
c_A^2=\frac{q_A}{\tau_A^2}-1\ .
\end{equation}

The unconditional service time $T$ for a randomly selected task is $T_H$ with probability $1-x$ and $T_A$ with probability $x$. Its first two moments are
\begin{align}
m(x;r)&=\mathbb E[T]=(1-x)\tau_H+x \tau_A\ ,\label{eq:m}\\
q(x;r)&=\mathbb E[T^2]=(1-x)\tau_H^2(1+c_H^2)+x \tau_A^2(1+c_A^2)=(1-x)\tau_H^2(1+c_H^2)+x q_A\ .\label{eq:q}
\end{align}

\subsection{Waiting time: a brief review}
\label{sec:waiting}

 We restate here two classical results that we used above. Substituting the mixed-system moments in Eq. \eqref{eq:m} and Eq. \eqref{eq:q} into Eq. \eqref{eq:PK} gives the M/G/1 mean waiting time $W_q$ for our model in Eq. \eqref{eq:Wq_model}.

\subsubsection{The Pollaczek--Khinchine formula}\label{sec:PKsetting}

Consider an \(M/G/1\) queue: a single server processes tasks one at a time, tasks arrive according to a Poisson process with rate \(\lambda\), and each task has a random service time \(S\). Here \(S\) means the amount of calendar time for which the server is occupied by one task. This is related to the human-attention requirement \(T\) from Section~\ref{sec:modelmain} as
follows: the unconditional service time \(T\) is measured in units of human-attention time. The
team supplies \(C\) units of human-attention capacity per unit of calendar
time. Therefore the corresponding calendar service time is $S=T/C$. Consequently, $\mathbb E[S]=\mathbb E[T]/C$ and 
$\mathbb E[S^2]=\mathbb E[T^2]/C^2$, where the average is taken over independent and identically distributed tasks, with arbitrary distribution. 

The utilisation rate is the average fraction of time for which the server is busy
\begin{equation}
\rho=\lambda\mathbb E[S]=\lambda\frac{\mathbb E[T]}{C}=\frac{\lambda m(x;r)}{C}\ .
\end{equation}

The queue is stable only if \(\rho<1\): average work arriving per unit time must be smaller than the server's processing capacity.

Under the stability condition \(\rho<1\), the Pollaczek--Khinchine formula \cite{pollaczek1930,khinchine1932} gives
\begin{equation}
W_q^{M/G/1}
=
\frac{\lambda \mathbb E[S^2]}{2(1-\rho)}
=
\frac{\rho}{1-\rho}\cdot \frac{1+c_s^2}{2}\cdot \mathbb E[S]\ ,
\label{eq:PK}
\end{equation}
where $c_s^2 = \Var(S)/\E[S]^2$ is the squared CV of service time.

The derivation, due originally to Khinchine, uses the PASTA property (Poisson Arrivals See Time Averages) and Little's law \cite{wolff1982pasta,little1961proof}; standard textbook treatments include \cite{kleinrock1975queueing,cohen1969}, and \cite{asmussen2003applied}. The economic content of Eq. \eqref{eq:PK} is that \emph{mean waiting time depends on the variance of service time, not only its mean}. Two service-time distributions with identical means, but different variances, will produce different waiting times under the same arrival process. As $\rho \to 1^-$, the waiting time diverges; the prefactor $(1 + c_s^2)/2$ multiplies the divergence, so high-variance systems break down faster as load increases.

\subsubsection{The Kingman approximation for G/G/1}\label{sec:Kingman}

The \(M/G/1\) formula assumes Poisson arrivals. This means that arrivals are random but have no extra clustering beyond the randomness implied by a Poisson process. Equivalently, the time between two consecutive arrivals has squared CV equal to one. We denote this quantity by $c_a^2$:
\begin{equation}
c_a^2=
\frac{\operatorname{Var}(\text{interarrival time})}
{\mathbb E[\text{interarrival time}]^2}=1\ .
\end{equation}
This assumption lets us use the Pollaczek--Khinchine formula in Section \ref{sec:waiting}.

This is a convenient assumption, but many real workflows are more clustered: tasks may arrive in bursts around deadlines, shift changes, product releases, or end-of-period reporting. To describe such cases, queueing theory uses the more general \(G/G/1\) model, where the first \(G\) means that the interarrival-time distribution is general rather than Poisson; the second \(G\) means that the service-time distribution is general; and the \(1\) means that there is one effective server or bottleneck. 

Let \(c_a^2\) denote the squared CV of interarrival times, and let \(c_s^2\) denote the squared CV of service times. Refs. \cite{kingman1961,kingman1962,kingman1970}  showed that, under suitable heavy-traffic conditions, the mean waiting time in a \(G/G/1\) queue is approximately

\begin{equation}
\label{eq:Kingman}
W_q^{\text{G/G/1}} \;\approx\; \frac{\rho}{1-\rho} \cdot \frac{c_a^2 + c_s^2}{2} \cdot \E[S]\ ,
\end{equation}
which is exact asymptotically as $\rho \to 1^-$. Equation~\eqref{eq:Kingman} is widely used as a robust operational approximation at moderate utilization, where it is known to be reasonably accurate \cite{kleinrock1975queueing,whitt1983}; for operational approximations based on squared coefficients of variation, see
\cite{whitt1983}; for a manufacturing-oriented treatment, see
\cite{hopp2001factory}. We shall refer to it as the Kingman approximation. For Poisson arrivals, $c_a^2 = 1$ and Eq. \eqref{eq:Kingman} reduces to Eq.\eqref{eq:PK}.

The Kingman approximation makes burstiness explicit: a bursty arrival process ($c_a^2 \gg 1$) inflates waiting times just as effectively as a high-variance service process. For applications of our model to specific workflows, $c_a^2$ should be measured directly from interarrival data.

\subsection{Derivation of the variance wedge boundary (Eqs. \eqref{eq:trap_CV} and \eqref{eq:design})}\label{app:variancetrap}

We recall that, for a pure manual system (\(x=0\)), the human-attention requirement is \(T_H\), with
\begin{equation}
\mathbb{E}[T_H]=\tau_H\ ,
\qquad
\operatorname{Var}(T_H)=\sigma_H^2\ ,
\qquad
c_H^2=\frac{\sigma_H^2}{\tau_H^2}\ .
\end{equation}
Its second raw moment is, therefore,
\begin{equation}
q_H
\coloneqq
\mathbb{E}[T_H^2]
=
\tau_H^2+\sigma_H^2
=
\tau_H^2(1+c_H^2)\ .
\end{equation}
Substituting \(m=\tau_H\) and \(q=q_H\) into the \(M/G/1\) waiting-time formula gives
\begin{equation}
W_H
=
\frac{\lambda q_H}{2C(C-\lambda\tau_H)}
=
\frac{\lambda \tau_H^2(1+c_H^2)}
{2C(C-\lambda\tau_H)}\ .
\end{equation}

For a pure AI-routed system (\(x=1\)), we define
\begin{equation}
\tau_A
\coloneqq
\mathbb{E}[T_A],
\qquad
q_A
\coloneqq
\mathbb{E}[T_A^2],
\qquad
c_A^2
=
\frac{q_A}{\tau_A^2}-1\Rightarrow 
q_A=\tau_A^2(1+c_A^2)\ .
\end{equation}
Substituting \(m=\tau_A\) and \(q=q_A\) into the same formula gives
\begin{equation}
W_A
=
\frac{\lambda q_A}{2C(C-\lambda\tau_A)}
=
\frac{\lambda \tau_A^2(1+c_A^2)}
{2C(C-\lambda\tau_A)}\ .
\end{equation}

Assuming both systems are stable, \(W_A<W_H\) if and only if
\begin{equation}
\frac{\tau_A^2(1+c_A^2)}{C-\lambda \tau_A}
<
\frac{\tau_H^2(1+c_H^2)}{C-\lambda\tau_H}\ .
\end{equation}
Dividing by \(\tau_H^2(1+c_H^2)\) and multiplying numerator and denominator by
\(1/C\) gives
\begin{equation}
\frac{1+c_A^2}{1+c_H^2}
<
\left(\frac{\tau_H}{\tau_A}\right)^2
\frac{1-\lambda \tau_A/C}{1-\lambda\tau_H/C}\ .
\end{equation}
Letting \(s=\tau_A/\tau_H\) and \(\rho_H=\lambda\tau_H/C\), this becomes
\begin{equation}
1+c_A^2
<
\frac{1}{s^2}
\frac{1-\rho_Hs}{1-\rho_H}
(1+c_H^2)\ ,
\end{equation}
so the boundary value is
\begin{equation}
c_{A,\max}^2(s,\rho_H)
=
\frac{1}{s^2}
\frac{1-\rho_Hs}{1-\rho_H}
(1+c_H^2)-1\ .
\end{equation}

\subsection{Proof of ``bang-bang'' solution in the case of partial adoption (Eq. \eqref{eq:bang_bang_x})}\label{sec:bangbang}

Again, we write $\tau_A:=m(1;r)$, $q_A:=q(1;r)$, $c_A^2:=q_A/{\tau_A}^2-1$ and $q_H:=\tau_H^2(1+c_H^2)$. We also write for simplicity $m(x):=m(x;r)$, $q(x):=q(x;r)$, and $W_q(x)=W_q(x;r)$.
Then
\begin{equation}
m(x)=\tau_H+x(\tau_A-\tau_H)\ ,
\end{equation}
and
\begin{equation}
q(x)=q_H+x(q_A-q_H)\ .
\end{equation}
Substituting into the \(M/G/1\) formula \eqref{eq:PK} gives
\begin{equation}
W_q(x)
=
\frac{\lambda q(x)}
{2C(C-\lambda m(x))}\ .
\end{equation}
Differentiating with respect to \(x\),
\begin{equation}
\frac{dW_q}{dx}
=
\frac{\lambda}{2C}
\frac{
(q_A-q_H)(C-\lambda m(x))
+
\lambda ~q(x)(\tau_A-\tau_H)
}
{(C-\lambda m(x))^2}\ .
\end{equation}
Since both \(m(x)\) and \(q(x)\) are linear in \(x\), the numerator simplifies to $
C(q_A-q_H)+\lambda(q_H\tau_A-q_A\tau_H)$,
which is independent of \(x\). Hence \(W_q(x)\) is monotone on the stable region.

Equivalently, the derivative is negative exactly when
\begin{equation}
\label{eq:marginal_x_same_boundary}
\frac{q_A}{C-\lambda \tau_A}
<
\frac{q_H}{C-\lambda\tau_H}\ .
\end{equation}
This is precisely the same inequality obtained by comparing the two pure systems,
\(W_A<W_H\), and is equivalent to the variance-wedge condition
\eqref{eq:trap_CV} after writing \(q_A=\tau_A^2(1+c_A^2)\) and
\(q_H=\tau_H^2(1+c_H^2)\). Thus, in the homogeneous baseline model, there is no
separate intermediate adoption threshold for waiting-time minimization: the same
boundary that decides whether full AI routing beats full manual handling also
decides whether increasing \(x\) locally raises or lowers waiting time at every
stable intermediate value of \(x\).

\subsection{Review threshold under congestion (Section \ref{sec:thresholdcong})}\label{sec:reviewthreshold}
We consider a draft with posterior error risk \(\pi(s)\), where review reduces the remaining error probability according to Eq.~\eqref{eq:p_r}, where for definiteness, we set here \(p_\infty=0\): this means that with enough review, all reducible errors can in principle be caught. 

Each unit of review effort, therefore, multiplies the reducible risk by \(e^{-\kappa r}\). Thus, in the signal-dependent problem, the residual probability that an error escapes after review is \(\pi(s)e^{-\kappa r}\) rather than the unconditional \(p(r)\) in Eq.~\eqref{eq:p_r}. The reviewer therefore chooses \(r\geq 0\) to minimize
\begin{equation}
\theta r + K\pi(s)e^{-\kappa r},
\label{eq:choosetominimise}
\end{equation}
where \(\theta>0\) is the congestion cost of using scarce reviewer time and \(K>0\) is the loss if an error escapes. The first term is the cost of spending time on review rather than moving on to other tasks in the queue; the second term is the expected cost of an escaped error after review.

 If an irreducible error component is retained, the escape probability becomes
\begin{equation}
\pi(s) [p_\infty+(1-p_\infty)e^{-\kappa r}]\ .
\end{equation}
The term \(K\pi(s)p_\infty\) is independent of \(r\), so the optimal review rule
is unchanged except that \(\kappa K\) is replaced by \(\kappa K(1-p_\infty)\).

The first-order condition for the optimal $\theta$ that minimizes \eqref{eq:choosetominimise}
\begin{equation}
\theta = \kappa K \pi(s) e^{-\kappa r}
\end{equation}
yields, after applying the non-negativity constraint \(r \geq 0\),
\begin{equation}
\label{eq:r_star}
r^\star(s; \theta)
=
\frac{1}{\kappa} \max\!\left\{0,\; \log\frac{\kappa K \pi(s)}{\theta}\right\}\ .
\end{equation}
Review is positive only when \(\pi(s)\) exceeds the threshold
\begin{equation}
\label{eq:pi_star}
\pi^\star(\theta)
=
\frac{\theta}{\kappa K}\ .
\end{equation}
If \(\pi(s)\leq\pi^\star(\theta)\), review is not worth the time: the draft looks safe enough relative to the cost of delaying other work. If \(\pi(s)>\pi^\star(\theta)\), the reviewer spends time checking it, and the amount of review increases logarithmically with the perceived risk \(\pi(s)\).


The earlier $r$-dependent moments \(\tau_A\) and \(q_A\) in Eq. \eqref{eq:ar1} and Eq. \eqref{eq:a2} were computed for a fixed review time \(r\). We now allow review time to depend on the draft's perceived risk. A task with signal \(s\) receives review time \(r^\star(s;\theta)\), so its mean and second moment are still given by the same formulas as in Equations \eqref{eq:ar1} and \eqref{eq:a2}, with \(r\) replaced by \(r^\star(s;\theta)\). To obtain the average AI-route moments for the whole workflow, we average these quantities over the distribution of signals \(\mu(s)\).

The resulting AI-route moments are
\begin{equation}
\tau_A(\theta)
=
\mathbb E_s[r^\star(s;\theta)]
+
\mu_R
\mathbb E_s\!\left[\pi(s)e^{-\kappa r^\star(s;\theta)}\right]\ ,
\label{eq:routemoment1}
\end{equation}
and
\begin{equation}
q_A(\theta)
=
\mathbb E_s\!\left[
r^\star(s;\theta)^2
+
2r^\star(s;\theta)\mu_R\pi(s)e^{-\kappa r^\star(s;\theta)}
+
\mu_{R,2}\pi(s)e^{-\kappa r^\star(s;\theta)}
\right]\ ,
\label{eq:routemoment2}
\end{equation}
where
\begin{equation}
\mathbb E_s[f(s)] := \int_{\bar S} f(s)\mu(s)\,ds\ .
\end{equation}

If a reviewer spends more time checking an AI draft, the average human time required per task rises. This pushes the system closer to capacity and increases waiting time for other tasks. We define \(\theta\) as the extra waiting cost created by a small increase in average human-attention demand. We close the model by imposing that the `right' congestion cost \(\theta\) must be equal to the marginal waiting cost created by one additional unit of mean human-attention demand:
\begin{equation}
\theta
:=
c_w\lambda
\frac{\partial W_q}{\partial \tau_A}
\bigg|_{\tau_A=\tau_A(\theta),\,q_A=q_A(\theta)}.
\label{thetaaswaitingtimerises}
\end{equation}
Here \(W_q\) is the mean waiting time in Eq.~\eqref{eq:Wq_model}. The only change is that the service-time moments now depend on \(\theta\), because the review rule \(r^\star(\cdot;\theta)\) determines the AI-route mean \(\tau_A(\theta)\) and second moment \(q_A(\theta)\) via Eqs. \eqref{eq:routemoment1} and \eqref{eq:routemoment2}. In words, \(\partial W_q/\partial \tau_A\) (keeping $q_A$ fixed) tells us how much average waiting time rises when the average human time per AI-routed task increases slightly. Multiplying by \(\lambda\) converts this into the total waiting-time increase across arriving tasks, and multiplying by \(c_w\) (where \(c_w>0\) measures how costly it is for a task to wait in the queue) converts waiting time into cost. Thus, \(\theta\) in Eq. \eqref{thetaaswaitingtimerises} is the queue-generated cost of using a little more reviewer time.

Therefore, equation
\eqref{thetaaswaitingtimerises} defines a fixed-point problem for \(\theta\), where a verification equilibrium is any value $\theta^\star$ such that 
\begin{equation}
\theta^\star=\Phi(\theta^\star)\ ,
\label{eq:verification_fixed_point}
\end{equation}
with $\Phi(\theta)$ the right-hand side of \eqref{thetaaswaitingtimerises}. Under fairly general assumptions, a fixed-point solution exists by Brouwer's theorem, although it may not be unique, as we saw in the example in Fig. \ref{fig:theta_fixed_point}. When several fixed points exist, we report the smallest stable solution. Since \(\theta\) is the congestion cost of reviewer time, this root corresponds to the least-congested self-consistent verification regime and to the branch naturally reached from a lightly loaded system. Higher roots describe more congested, high-threshold regimes; selecting among them would require an explicit dynamic adjustment rule, which is beyond the scope of the static model.

From \eqref{thetaaswaitingtimerises}, an exogenous increase in \(\lambda\) or \(c_w\) raises the right-hand side at fixed \(\theta\), and provided the fixed-point map is well behaved, in particular monotone in load and bounded away from saturation, the equilibrium \(\theta^\star\) rises with both. A higher \(\theta^\star\) also raises \(\pi^\star\), which by \eqref{eq:r_star} reduces review effort on every draft: it is set to zero for more drafts, and lowered for drafts above the threshold. Note however that monotonicity is plausible given that congestion makes attention more valuable, but it is not automatic because \(\tau_A(\theta)\) and \(q_A(\theta)\) both move endogenously.

{\bf Acknowledgments:} P.V. acknowledges support from UKRI FLF Scheme (No. MR/X023028/1).


\end{document}